\documentclass[twocolumn,showpacs,preprintnumbers,amsmath,amssymb]{revtex4}

\usepackage{graphicx}

\usepackage{dcolumn}
\usepackage{bm}

\begin{document}


\title{Predicted Ultrafast Single Qubit Operations in Semiconductor Quantum Dots}

\author{C. E. Pryor}
 \email{craig-pryor@uiowa.edu}
\author{M. E. Flatt\'e}%
\affiliation{%
Optical Science and Technology Center and Department of Physics and Astronomy, University of Iowa, Iowa City, Iowa, 52242, USA}%

\date{\today}

\begin{abstract}
Several recently proposed implementations of scalable quantum computation rely on the ability to manipulate the spin polarization of individual electrons in semiconductors. The most rapid single-spin-manipulation technique to date relies on the generation of an effective magnetic field via a spin-sensitive optical Stark effect. This approach has been used to split spin states in colloidal CdSe quantum dots  and to manipulate ensembles of spins in ZnMnSe quantum wells with femtosecond optical pulses.  
Here we report that the process will produce a coherent rotation of spin in quantum dots containing a single electron.
The calculated  magnitude of the effective magnetic field depends on the dot bandgap and the strain.  We predict that in InAs/InP dots, for reasonable experimental parameters, the magnitude of the rotation is sufficient and the intrinsic error is low enough for them to serve as elements of a quantum dot based quantum computer.
\end{abstract}


\maketitle


Scalable proposals for quantum computation\cite{Burkard.book.2002}  require well-defined individual qubits that can be manipulated individually (single-qubit gates) and also can controllably interact with each other (two-qubit gates)\cite{Barenco.pra.1995}. In solid-state systems often the two-qubit gates appear easier to realize, e.g.  by the exchange interaction between electrons\cite{Loss.pra.1998}. The manipulation of single qubits has been perceived as more difficult; this motivated the proposal of Òall-exchange-basedÓ quantum computation\cite{DiVincenzo.nature.2000}, which requires only two-qubit (exchange) gates.   ÒAll-exchange-basedÓ  computations require a large number of gate operations, although recent work indicates substantial reductions are possible\cite{Levy.prl.2002}."

Here we predict that a spin-AC Stark effect (1) produces coherent rotations of electron spin in quantum dots, (2) the rotation angle can exceed $\pi$ for reasonable experimental parameters, and (3) the error rates are tolerable for quantum computation.  The spin-splittings of states in CdSe quantum dots\cite{Gupta.prb.2001,Gupta.prb.2002} and coherent manipulation of ensembles of spins in ZnMnSe quantum wells\cite{Gupta.science.2001}  has been demonstrated experimentally.  Theoretical considerations of the AC Stark effect \cite{Cohen-Tannoudji.pra.1972, Combescot.prl.1988,Cohen-Tannoudji.book.1998}  have focused to date only on non-spin-selective shifts of energy levels.  Here we consider a spin-selective AC Stark effect whereby a quantum dot is illuminated with a single intense pulse of circularly-polarized nonresonant light (Fig. 1a). Such a pulse shifts the energies of dot states, and due to differing transition matrix elements the two spin states are shifted differently. Hence this pulse produces a splitting of the two lowest energy conduction states. The splitting of these two states (spin-up and spin-down) can be viewed as an optically-induced pseudo-magnetic field ( $B_{eff}$ ) acting on the state pair. Direct application of this approach to quantum computing is clear, for one well-known physical realization of a single qubit operation is a magnetic field applied to a spin for a definite period of time. We find that the size of the effective field in certain dot systems, for reasonable experimental conditions, exceeds the magnitude required for 180 degree rotation of the spin polarization ($\pi$-pulse).   Furthermore we evaluate the error rates for spin manipulation via this process under reasonable conditions to be smaller than $10^{-6}$, which is within the tolerance for quantum computation.

\begin{figure}
\includegraphics [width=8.5cm]{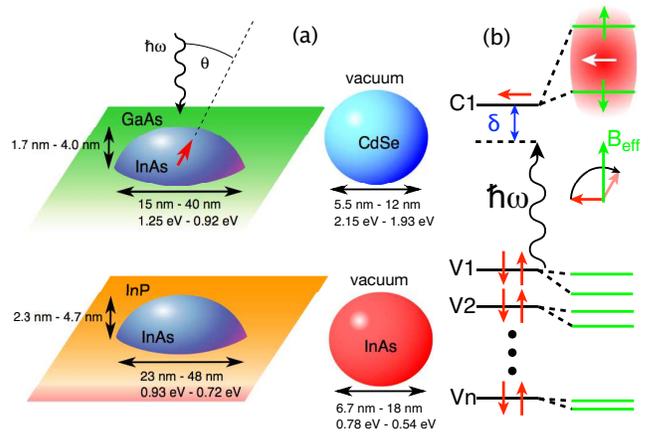}
\caption{\label{fig:1} a) SAQD and CQD quantum dot geometries.  For SAQDs, the incident light is taken to be incident along the growth direction.   b)  Level diagram for a  quantum dot with one electron in the lowest conduction band state, perturbed by incident light detuned below the bandgap.  The electron spin precesses in response the the effective magnetic field $B_{eff}$ corresponding to the spin splitting.  }
\end{figure}

There are a wide variety of quantum dot systems from which to choose, with  various properties affecting the size of the Stark splitting.  Of particular interest are the level spacings and oscillator strengths, both of which are affected by material parameters and the size of the dots.  We will examine Stark splittings in CdSe and InAs colloidal quantum dots (CQD), as well as in InAs/GaAs and InAs/InP self assembled quantum dots (SAQD).  This set, while not exhaustive, spans a broad range of parameters by including unstrained spherical and strained lens-like dots with widely varying bandgaps.

The AC Stark effect with unpolarized light is a nonlinear effect whereby light with photon energy tuned near to the absorption transition between two states induces a level repulsion between the two states\cite{Cohen-Tannoudji.pra.1972, Cohen-Tannoudji.book.1998}. In perturbation theory\cite{Combescot.prl.1988} this level repulsion depends on $I f/\delta$, where $I$ is the light intensity, $f$ is the oscillator strength of the transition, and $\delta$  is the detuning of the photon energy from the transition energy (see Fig. 1b). When circularly polarized light illuminates a transition from the first valence state pair to the first conduction state pair in a crystal with spin-orbit interaction the oscillator strengths of the transitions differ considerably; this is the source of the effective spin splitting of the conduction state pair. 

\begin{table}
\begin{ruledtabular}
\begin{tabular}{ccccc}
Parameter  &  InAs  &  GaAs  &  InP  &  CdSe\\
\hline
$E_g$ (eV)& 0.418 & 1.519 & 1.424 & 1.84\\
$\Delta_{so}$ (eV) & 0.38 & 0.341 & 0.11  &0.42\\
VBO (eV)& 0.085 & 0.0 & -0.045 & -\\

$m^*_e$ &   -   &   -   &   -     &0.119\\
$\gamma_1$ & 19.67 & 6.98 & 4.95   &2.52\\
$\gamma_2$ & 8.37 & 2.25 & 1.65    & 0.65\\
$\gamma_3$ & 9.29 & 2.88 & 2.35   & 0.95\\
$E_P$ (eV) & 22.2 & 22.7 & 20.4   & 17.4 \\

$a_c$ (eV)& -6.67 & -9.55 & -7.33 &-\\
$a_v$ (eV)& 1.67 & 0.95 & 0.73 &-\\
$b$ (eV)    & -1.8 & -2.0 & -2.0    &-\\
$d$ (eV)    & -3.6 & -5.4 & -5.0    &-\\

$a_{latt}$ (nm) & 0.6058 & 0.5653 & 0.5869& -\\
$C_{xxxx}$ (GPa)& 832.9 & 1211 & 1022 &-\\
$C_{xxyy}$ (GPa)& 452.6 & 548 & 576  &-\\
$C_{xyxy}$ (GPa)& 395.9 & 604 & 460  &-\\

\end{tabular}
\end{ruledtabular}
\caption{\label{tab:table1}
Material Parameters used in the calculations\cite{Landolt.1982}. The valence band offset (VBO) is the unstrained valence energy, determining the band alignment in SAQDs.  For CQDs infinite barriers were used for the vacuum. }
\end{table}

Optical Stark shifts were calculated non-perturbatively using a restricted basis of quantum dot wave functions calculated with $\rm k \cdot p$ theory in the envelope approximation.  The wave functions were calculated numerically on a real-space cubic grid using the Lanczos algorithm\cite{pryor.prb.1997}. This method has been used previously to calculate a variety of  electronic properties of strained QDs.  The grid spacing was chosen to match the lattice constant of the barrier material in the SAQDs, and the dot material itself for the CQDs.  The number of grid sites varied from $10^4 - 2\times 10^5$ depending on the size of the QD.  For SAQDs the strain arising from lattice mismatch was calculated by doing a conjugate gradient  minimization of the continuum elasticity strain energy. The strain was then used as input to an eight-band strain-dependent Hamiltonian, which was diagonalized as described above.  Because of the larger bandgap and lack of strain, the CdSe calculations were performed with a single-band model for the conduction band, and a four-band model for the valence band.  The CdSe calculations were done for the zincblende form. The material parameters are given in Table I.

Calculation of the energy shifts under illumination was done by constructing a restricted set of Fock states for the combined electron-photon system. Eight valence states and the lowest conduction doublet were used for the quantum dot states. The states in the basis for the electron-photon system were $| 1,0;1,1,1,1,1,1,1,1;N_\gamma \rangle$, $|0,1,1,1,1,1,1,1,1,1; N_\gamma \rangle$, $|1,1; 1,1,\dots 0, \dots , (N-1)_\gamma \rangle $
 for a total of 10 states. For the first two Fock states all valence states and one conduction state are occupied, and there are $N_\gamma$  photons in the photon field. For the other eight Fock states, both conduction states are occupied, only one valence state is unoccupied, and there is one less photon than in the first two states. Inclusion of additional states did not alter the results. 
In this restricted basis the Hamiltonian for the coupled electron-photon system is given by 

\begin{eqnarray}
H  &=&
\begin{pmatrix}
E_{c1} - \hbar \omega  & 0                                      & d_{11} & d_{21}&\dots \\
0                                       & E_{c2} - \hbar \omega & d_{12}   & d_{22}&\dots \\
d^*_{11}                         & d^*_{12}                         & E_{v1} & 0           & dots \\
d^*_{21}                         & d^*_{22}                         &0            & E_{v2} & \dots \\
\vdots                              & \vdots                             &\vdots    & \vdots    & \\   &  
\end{pmatrix}\\
d_{ij} & =& {e E_0 \over 2 m_e \omega } \langle \psi_{vi} | \vec P \cdot \vec \epsilon | \psi_{cj} \rangle \\
E_{c1}  &=& E_{c2}, E_{v1} = E_{v2},\dots
\end{eqnarray}
where $\vec P $  is the momentum operator, $\vec \epsilon $  is the polarization vector of the incident light with electric field $E_0$, $\psi_{vi}$ and $\psi_{cj}$ are the valence and conduction band states respectively, and $d_{ij}$ are the dipole matrix elements.  Coherent states for the light field can then be naturally constructed from the mixed electron-photon states obtained above.

 In Fig 2a and 2b are shown the energies of the conduction states for two representative dot systems:  an InAs/GaAs SAQD and a CdSe  CQD. In both cases,the splittings increase as the detuning approaches zero, and change sign when the detuning changes sign. 
The first notable feature is the difference in spin selectivity, which can be traced to the differing strain and shape.  For CdSe both spin states are shifted, but by different amounts, while for InAs/GaAs only one spin state is shifted.  Because the CdSe dot has approximate spherical symmetry and is unstrained, the highest valence states are closely spaced (  $< 1 ~\rm meV$), with comparable amounts of heavy hole (HH) and light hole (LH) character.  As a result,  both the HH and LH transitions  contribute to the Stark shift, and both spin directions for the conduction state are shifted.  The difference between the shifts reflects the difference in oscillator strengths for HH and LH transitions, whose ratio of 2:1 is approximately the ratio found in bulk band-to-band transitions, 3:1.    In contrast, for InAs/GaAs dots the highest valence state is almost entirely HH, and separated from the next valence state by several tens of meV (depending on size).  Hence, the Stark shift is dominated by the (doubly degenerate) highest valence state which gives a Stark shift for only one spin direction of the conduction electron. 

We now examine the effect of dot size on the spin splitting.  Shown in Fig. 3 are the splittings for detunings from 30 meV - 70 meV for the four  systems under consideration. Since we are ultimately interested in manipulating spins through the effective magnetic field, it is useful to consider  the precession angle associated with a light pulse of duration $\delta t$ , given by $\theta_s = \Delta E  \delta t / \hbar$  where $\Delta E$ is the Stark spin splitting between up and down conduction states.  In the results that follow, we give Stark splittings in both meV, and the corresponding $\theta_s$  for a 200 fs pulse with power density $10^9 \rm W/cm^2$, which we refer to as a reference pulse.  The $\sim 2$meV spin splitting seen in CdSe CQD's agrees with that measured experimentally \cite{Gupta.prb.2001}.

\begin{figure}
\includegraphics  [width=8.5cm] {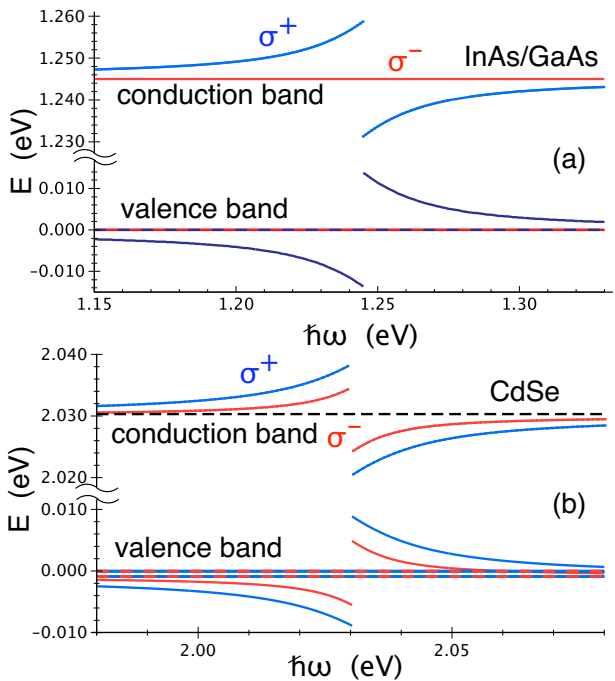}
\caption{\label{fig:2}  a) Energy levels for an InAs/GaAs dot with 1.245 eV bandgap.  The incident power density is $10^9~ \rm  W/cm^2$ and the conduction band electron has spin parallel to the direction of the incident light.    b)  CdSe dot with 2.030 eV bandgap.  }
\end{figure}
\begin{figure}
\includegraphics  [width=8.5cm]{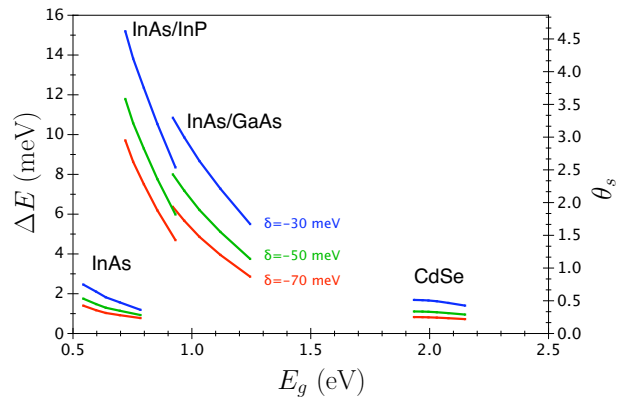}
\caption{\label{fig:2} Stark splittings as a function bandgap and detuning for InAs, InAs/InP, InAs/GaAs, and CdSe quantum dots.  The angle $\theta_s = \Delta E \delta t / \hbar $  is for an incident 200 fs pulse with $10^9~ \rm W/cm^2$.  }
\end{figure}

\begin{figure}
\includegraphics  [width=8.5cm] {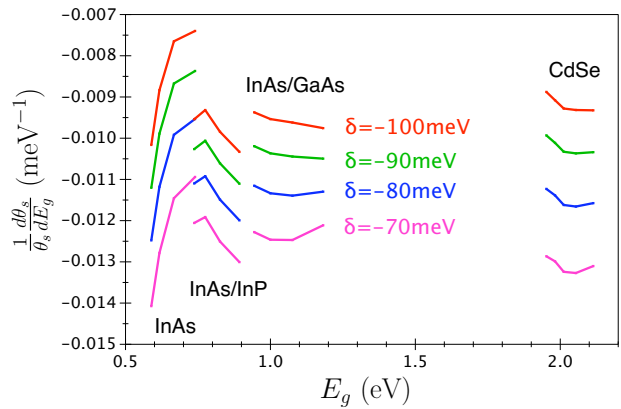}
\caption{\label{fig:2} Sensitivity of $\theta_s$  to variations in the bandgap as a function of bandgap and detuning, estimated using a finite difference approximation.  Note that larger detunings than in Fig. 2 were used so there would be values of $\theta_s$ at fixed values of $\hbar \omega$  for every bandgap value.  $\theta_s$ is computed assuming a reference pulse  }
\end{figure}

The trend in Fig. 3 is surprising; for a given material system larger dots have larger splittings, in spite of the fact that oscillator strengths decrease with increasing dot size.  The reason is that the the interaction term in the Hamiltonian is proportional to $1/\omega$.  For a fixed detuning, the increase in $1/\omega$  for larger dots dominates the decrease in the dipole matrix element.  

The importance of the bandgap suggests selecting the dot material to minimize the gap.  One possibility is to use a small bandgap material, such as InAs, but to avoid the bandgap-increasing effect of strain by using an InAs CQD. Confinement  increases the bandgap over that of bulk InAs, but for typical CQD sizes $E_g < 1~ \rm eV$, which is substantially less than $E_g \approx 2 ~\rm eV$ for CdSe CQDs.

Fig. 3 confirms the expectation that InAs CQDs have larger splittings than CdSe CQDs over most of the range of sizes, but the increase in splitting is at most a factor of 1.5.  While InAs CQDs have larger splittings than CdSe, they are still a factor of 5 smaller than for InAs/GaAs.  In spite of a smaller bandgap, the lack of strain in InAs CQDÕs decreases the HH/LH splitting so much that the Stark splitting is substantially decreased as well.

An alternative to the complete elimination of strain is to choose a substrate material that has a lattice constant closer to InAs.  InAs/InP is an excellent candidate, with a lattice mismatch of 4\% (as opposed to 7\% for InAs/GaAs), and a edge-to-edge bandgap of approximately $ 0.5 ~\rm eV$\cite{Pettersson.prb.2000, Holm.jap.2002}.   The highest valence state is still predominantly HH, and the first excited valence state ranges from 10 meV - 30 meV below, depending on size.  In addition, InAs/InP has a bandgap at the technologically important $1.55 ~\mu \rm m$.  Fig. 3 shows that the splittings in InAs/InP are substantially larger than those for InAs/GaAs.  Some of the improvement is simply due to the smaller bandgap and the factor of $1/\omega$  in the interaction as discussed earlier. However, the discontinuity in the curves indicates the improvement due to the increase in strain and the resulting increased HH/LH splitting.  At our standard illumination of $10^9 ~\rm W/cm^2$ with a 200 fs pulse width, a $\pi$-rotation is possible at even large detunings, $\delta = -70 \rm meV$. 

We now consider errors in $\theta_s$  due to the (unavoidable) variations in the bandgap of the dots.  For measurements on an ensemble of dots, inhomogeneities in the dot size will yield a different rotation angle for different dots.  More relevant for a quantum computer, in which individual dots will be selected, the finite line width of the states will cause some uncertainty in $\theta_s$.  To address this question, we have calculated numerical estimates of  
\begin{eqnarray} 
{ 1 \over \Delta E} { \partial \Delta E(E_g,\hbar \omega) \over \partial E_g} = {1 \over \theta_s} {\partial \theta_s (E_g, \hbar \omega ) \over \partial E_g },
\end{eqnarray}
as shown in  Fig. 4.  For the materials and sizes considered this quantity varies from  $-0.008~ \rm meV^{-1}$  to $-0.014~ \rm  meV^{-1}$, depending on  detuning.  The results of Fig. 4 may be used to estimate the uncertainty in the rotation angle, $\Delta \theta_s $.  For example, an ensemble measurement of InAs/InP dots with an inhomogeneous linewidth of 50 meV would give $\Delta \theta_s \approx 0.5$ for a $\pi$-pulse, which should be sufficient to permit observation of a spin echo.  For a single dot with a linewidth of 0.1 meV, $\Delta \theta_s \approx 0.001$, corresponding to a bit error rate of  $10^{-6}$.  This is sufficiently small for error correction algorithms to apply\cite{Preskill.book.1997}  (desirable error thresholds have been estimated\cite{Preskill.book.1997} at $10^{-5} - 10^{-6}$).

Errors in $\theta_s$ may also arise from shot noise in the laser pulse.  The electric field $E_0$ appearing in Eq. 2 has some uncertainty due to variations in the number of photons in the pulse.  Assuming the laser is focused to $ 1 ~ \mu \rm m^2 $ with a photon energy of 1 eV,  a reference pulse contains  approximately $ 10^7$ photons.  The uncertainty in the incident intensity is  $\Delta I \approx I / \sqrt N_\gamma $, giving $\Delta \theta_s \approx 2 \pi / \sqrt N_\gamma \approx 10^{-3}$.  The corresponding  bit error rate is approximately $10^{-6}$, which is still acceptably small for error correction\cite{Preskill.book.1997}.  This estimate gives a lower limit on the error rate since it neglects additional sources of laser noise that would increase the error rate.

We conclude that use of the spin-AC Stark effect is a viable approach to single qubit manipulation in quantum dots.  The magnitude of the Stark splitting in SAQDs is 5-10 times larger  than in CQDs due to the strain-induced HH/LH splitting.  The Stark splitting also increases with decreasing bandgap, though the effect is smaller that of strain.  Because strain increases both the HH/LH splitting and the bandgap, finding the optimal system involves a seeking a system with sufficient strain to induce HH/LH splitting, but not so much as to make the bandgap too large.  We propose as a candidate system InA/InP dots, which are strained, but have a relatively small bandgap of 0.7-0.9 eV.  This energy range includes $1.55 ~\mu \rm m$, which may be important for interconnection within a quantum computer, or for quantum communication applications.  We find that for such dots, $\pi$-pulses may be obtained for experimentally realistic pulses ($10^9 ~\rm W/cm^2$, 200 ~fs) and  detunings (-70 meV).  For typical inhomogeneous broadening (50 meV) the variation in  rotation angle is  $\Delta \theta_s \approx 0.5$.  For a single dot with an intrinsic linewidth of 0.1 meV we estimate the bit error rate to be on the order of $10^{-6}$, and we estimate the bit error rate due to laser shot noise is also on the order of $10^{-6}$.  These error rates are sufficiently low for quantum error correction.

We wish to thank D. D. Awschalom and J. A. Gupta for enlightening discussions.  The work was  supported by the Army Research Office MURI  DAAD19-01-1-0541

\bibliography{physRevStyleJNames,semiconductor}
\end{document}